\documentclass[]{spie}

\usepackage{amsmath,amsfonts,amssymb}
\usepackage{graphicx}
\usepackage[colorlinks=true, allcolors=blue]{hyperref}

\usepackage{comment}

\title{SRA vs histograms in the noise analysis of SPAD}

\author[1,2]{Nikolay S. Perminov}
\author[1]{Maksim A. Smirnov}
\author[3]{Raoul R. Nigmatullin}
\author[1]{Anvar A. Talipov}
\author[1,2,*]{Sergey A. Moiseev}

\affil[1]{Kazan Quantum Center, Kazan National Research Technical University n.a. A.N.Tupolev-KAI, 10 K. Marx, Kazan 420111, Russia}
\affil[2]{Zavoisky Physical-Technical Institute of the Russian Academy of Sciences, 10/7 Sibirsky Tract, Kazan 420029, Russia}
\affil[3]{Department of Radioelectronics and Information-Measuring Technique, Kazan National Research Technical University n.a. A.N.Tupolev-KAI, 10 K. Marx, Kazan 420111, Russia}

\authorinfo{* s.a.moiseev@kazanqc.org}

\begin{document} 
\maketitle

\begin{abstract}
A comparative analysis of the method of histograms and the sequence of the ranged am-plitudes (SRA) for statistical parametrization of the operation regime of a single-photon avalanche photodetector is carried out. It is shown that the SRA method contains all the information, which can be obtained using the method of histograms, and also allows to give a quick robust description of the dark counts of the device for a short noise sample of $\sim10^3$ points, what open the way for the introduction of SRA approach into software of a high-sensitivity photodetectors.
\end{abstract}

\keywords{SRA, SPAD, noise, noninvasive, discrete statistics}

\section{Introduction}
Single-photon detectors based on avalanche photodiodes (APD) are actively used in locating systems, quantum optics and communications \cite{Zhang2015}.
However, the work of the APD is greatly complicated by dark noise, some of which (afterpulsing counts) can hardly be eliminated in practice \cite{Cova1991}. 
A number of approaches are widely used to characterize the operation of the APD \cite{Zhang2015}, the most significant of which is based on the construction of a histogram reduced from time intervals between counts.
At the same time, the unreduced set of such intervals contains a complete (in contrast to the histogram) information about the noise, necessary for the practical use of the detector.

\noindent
In this work, a comparison is made between the histogram method and the sequences of the ranged amplitudes (SRA) \cite{Nigmatullin2003} for quantitative parameterization the time intervals between dark counts of the APD.
The SRA method has already demonstrated its effectiveness in solving a wide range of problems of discrete statistics \cite{Baleanu2010}.
We show that the SRA fully includes the histogram method, and also allows a much quicker characterization of the operating regime of the detector than histograms, using a smaller noise sample.

\section{SRA method}
The ranked (in descending order) sequence of time intervals between the photodetector samples $\{x_k\}$ ($k=\overline{1,N}$) forms SRA $\{s_n\}$, where the index $n$ is the index in SRA, and $N$ is the sample size \cite{Nigmatullin2003}.
The SRA is a noninvasive (without loss of information) quantitative characteristic of the sample \cite{Baleanu2010} and is related to the distribution function $F(x_n,N)$ by the approximate relation \cite{Nigmatullin2008,Smirnov2017}:
\begin{align}\label{gen_eq}
F(s_n,N)\cong(N+1-n(s_n))/N.
\end{align}
The histogram by its definition approximates the probability density $\rho(x)=dF/dx$ with increasing $N$, remaining an invasive and non-smooth function (the smoothness increases with increasing N), wich depending on the method of partitioning the sample data.
Noises of light sources and dark noises corresponding to Poisson processes [1] have probability density $\rho(x)$ (for time intervals "$x$" between nearest samples), described by the distribution $dF/dx=\rho(x)=\lambda e^{-\lambda x}$, where $\lambda$ is the average frequency of the registered counts.
For Poisson processes, the SRA is described by the formula \cite{Smirnov2017} $s_n=\lambda^{-1} \operatorname{Ln}⁡(N/(n-1))$, which has one free parameter $\lambda$. For an arbitrary case, the function $F(x,N)$ can be found from the general expression (\ref{gen_eq}).

\section{SRA vs histograms}
A quantitative analysis of the stability of the method with a change in the sample size can be made on the basis of the coefficient of determination \cite{Baleanu2010} $R^2$ or the normalized measure of coincidence.
For Q subsamples $\{x_{q,k}\}$ of length $N$, where $q=\overline{1,Q}$, $k=\overline{1,N}$, $\{y_k\}=Q^{-1}\sum_{q=1}^Qx_{q,k}$ is the averaged sample, $y=N^{-1}\sum_{k=1}^Ny_k$ is the full average, the relative quadratic ($R^2$-like) deviation factor of the form
\begin{align}\label{deviation}
\varepsilon(N)=\left. Q^{-1}\sum_{k=1}^N\sum_{q=1}^Q(x_{q,k}-y_k)^2\middle/ \sum_{k=1}^N(y_k-y)^2 \right.,
\end{align}
that allows us to compare the length of the SRA set $\operatorname{SRA}[\{x_{q,k}\}]$ of length $N$ with the histogram set $\operatorname{Hist}[\{x_{q,k}\}]$ ($m=\overline{1,N_h}$) of the optimal length \cite{Orlov2013} $N_h(N)=[4(3(N-1)^2/4)^ {1/5}] \operatorname{div} [1]$ (the Mann-Wald criterion).
It is obvious that $\operatorname{Hist}[\{x_{q,k}\}]=\operatorname{Hist}[\operatorname{SRA}[\{x_{q,k}\}]]$ and the histogram is a function only from the SRA (the SRA fully includes the histograms). This fact suggests that it is impossible to obtain more information from the histograms in principle (by definition) than from the SRA. Conversely, SRA contains all information about the histogram (lossless).

\begin{figure}[htb]
\begin{center}
\includegraphics[width=0.7\textwidth]{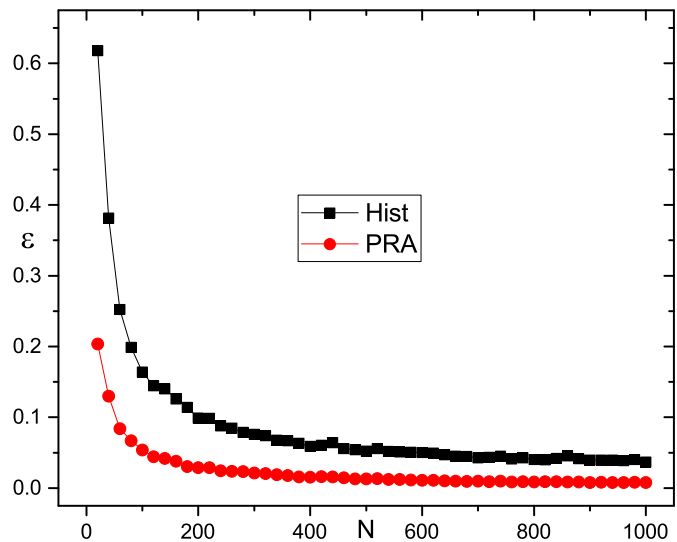}
\end{center}
\caption{
Dependence of the deviation $\varepsilon$ from $N$ for SRA (SRA - red circles) and histograms (Hist - black squares).}
\label{fig1}
\end{figure}

\noindent
Below, as an example of noise samples, we use the intervals between dark counts of a single-photon detector. For the efficiency of the detector ID210 (ID QUANTIQUE) 15\% and the dead time 24$\mu$s, a sample of a total length of $10^5$ points was experimentally obtained. From this sample, a set of $Q=100$ subsamples of length $N=20j$ ($j=\overline{1,50}$) was formed. Calculating the deviation $\varepsilon$ from (\ref{deviation}) for each $N$, in the case of the SRA and the histograms corresponding to the same subsamples, we obtain the graphs shown in Fig. \ref{fig1}.
As can be seen from Fig. \ref{fig1}, the deviation value $\varepsilon$ is much smaller for the SRA method than for the histograms for any values of $N$: $\varepsilon(10^3)=0.0369$ for the histogram and $\varepsilon(10^3)=0.0078$ for the SRA, that is, $\varepsilon$ is 4.7 times smaller, respectively. A large qualitative difference is also observed for other ways of selecting the partitioning parameter $N_h(N)$ because the histogram even with the maximum partitioning parameter $N_h(N)=N/10$ (see the criteria in \cite{Orlov2013}) still reduces the amount of data in statistical analysis from $N$ to $N/10$ points, which causes the principal invasiveness of the histogram method.

\begin{figure}[htb]
\begin{center}
\includegraphics[width=0.7\textwidth]{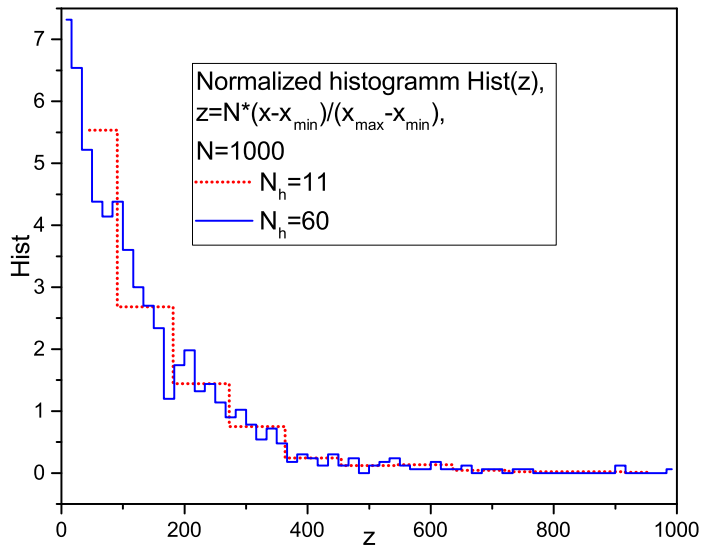}
\end{center}
\caption{
The normalized histograms $\operatorname{Hist}(z)$, $z=N(x-x_{min})/(x_{max}-x_{min})$, for the first subsample $\{x_k\}$ from Fig. \ref{fig1} for $N_h=11$ (red dotted curve, Sturges' criterion) and $N_h=60$ (blue curve, Mann-Wald criterion) (more optimal partitioning).}
\label{fig2}
\end{figure}

\noindent
A significant drawback of the histograms, in comparison with the SRA, is their dependence on $N_h(N)$ (see Fig. \ref{fig2}).
In this case, the choice of the optimal value of $N_h(N)$ is not determined by universal criteria and is a separate problem for each specific physical problem [6]. For a sample with $N=10^3$, corresponding to Fig. \ref{fig2}, and the theoretical fitting curve $\rho=\lambda e^{-\lambda x}$ for the Poisson processes described above, we have the deviation measure $1-R^2=0.0125$ for the optimal normalized to the average histogram ($N_h = 60$, Fig. \ref{fig2}) and $1-R^2=0.0043$ for the normalized to the average SRA ($R^2$ is the coefficient of determination \cite{Baleanu2010}).
That is, the fraction of errors in the fitting problem when using the SRA is approximately 3 times less. Another advantage of the SRA compared to the histograms \cite{Nigmatullin2003,Baleanu2010} is the gain in speed, which opens up important prospects for the introduction of SRA into the software of a high-sensitivity measurement technology.

\section{CONCLUSION}
The quantitative comparison of histograms and SRA demonstrates the significant practical advantages of the latter method for rapid statistical analysis of the noise of single-photon detectors due to the possibility of using short noise samples.
It is shown that the SRA method contains all the information that can be obtained using the histogram method.
Due to noninvasiveness, the SRA method is also applicable in the area of identification of various sources of a signal \cite{Baleanu2010,Spitsyn2016,Umnov2016} and noise \cite{Smirnov2017}.
The demonstrated advantages show the prospects of using the SRA approach for high-precision characterization of the noise of single-photon detectors, which is necessary for the realization of optical quantum computations and quantum communications.

\section*{ACKNOWLEDGMENTS}
Research of noise in the area of photonics and quantum technologies is financially supported by a grant of the Government of the Russian Federation, project No. 14.Z50.31.0040, February 17, 2017 (theoretical part). The work is also partially financially supported by the RFBR grant No. 16-32-60054 mol$\_$a$\_$dk (experimental part).

\bibliography{SRA_H}

\begin{thebibliography}{1}

\bibitem{Zhang2015}
Zhang, J., Itzler, M.~A., Zbinden, H., and Pan, J.-W., ``Advances in
  {InGaAs/InP} single-photon detector systems for quantum communication,'' {\em
  Light: Science {\&} Applications}~{\bf 4},  e286 (May 2015).

\bibitem{Cova1991}
Cova, S., Lacaita, A., and Ripamonti, G., ``Trapping phenomena in avalanche
  photodiodes on nanosecond scale,'' {\em IEEE Electron device letters}~{\bf
  12}(12),  685--687 (1991).

\bibitem{Nigmatullin2003}
Nigmatullin, R. and Smith, G., ``Fluctuation-noise spectroscopy and a
  “universal” fitting function of amplitudes of random sequences,'' {\em
  Physica A: Statistical Mechanics and its Applications}~{\bf 320},  291--317
  (2003).

\bibitem{Baleanu2010}
Baleanu, D., Guvenc, Z., and Machado, J.,  [{\em New Trends in Nanotechnology
  and Fractional Calculus Applications}{\nolinebreak\hspace{0.1em}]}, Springer
  Netherlands (2010).

\bibitem{Nigmatullin2008}
Nigmatullin, R., ``Strongly correlated variables and existence of a universal
  distribution function for relative fluctuations,'' {\em Physics of Wave
  Phenomena}~{\bf 16}(2),  119--145 (2008).

\bibitem{Smirnov2017}
Smirnov, M., Perminov, N., Nigmatullin, R., Talipov, A., and Moiseev, S.,
  ``Sequences of the ranged amplitudes as a universal method for fast
  noninvasive characterization of spad dark counts,'' {\em arXiv preprint
  arXiv:1708.07415}  (2017).

\bibitem{Orlov2013}
Orlov, Y.~N., ``Optimal histogram interval for non-stationary time-series
  distribution function density estimation,'' {\em Preprints of the Keldysh
  Institute of Applied Mathematics} ,  14--26 (2013).

\bibitem{Spitsyn2016}
Spitsyn, V.~G., Bolotova, Y.~A., Phan, N.~H., and Bui, T. T.~T., ``Using a haar
  wavelet transform, principal component analysis and neural networks for ocr
  in the presence of impulse noise,'' {\em Computer Optics}~{\bf 40}(2),
  249--257 (2016).

\bibitem{Umnov2016}
Umnov, A.~V. and Krylov, A.~S., ``Research of sparse representation method for
  ringing suppression,'' {\em Computer Optics}~{\bf 40}(6),  895--903 (2016).

\end{thebibliography}
\bibliographystyle{spiebib}

\end{document}